\documentclass[12pt]{article}
\input{epsf}
\def\bs{\vspace{1cm}}

\font\grande=cmr10 scaled \magstep4

\newcommand{\beq}{\begin{equation}}
\newcommand{\eeq}{\end{equation}}
\newcommand{\bea}{\begin{eqnarray}}
\newcommand{\eea}{\end{eqnarray}}

\def\laq{\raise 0.4 ex \hbox{$<$}\kern -0.8 em\lower 0.62 ex\hbox{$\sim$}}
\def\gaq{\raise 0.4 ex \hbox{$>$}\kern -0.7 em\lower 0.62 ex\hbox{$\sim$}}
\def\vev#1{\langle {#1}\rangle}

\def\frac#1#2{{\textstyle{{#1}\over {#2}}}}

\def\lsim{\mathrel{\rlap{\lower4pt\hbox{\hskip1pt$\sim$}}
    \raise1pt\hbox{$<$}}}
\def\gsim{\mathrel{\rlap{\lower4pt\hbox{\hskip1pt$\sim$}}
    \raise1pt\hbox{$>$}}}
\def\sqr#1#2{{\vcenter{\vbox{\hrule height.#2pt
         \hbox{\vrule width.#2pt height#1pt \kern#1pt
         \vrule width.#2pt}
         \hrule height.#2pt}}}}

\def\AJ{{\it Astrophys. J.} }

\def\GRG{{\it Gen. Relativity and Gravitation} }

\def\MNRAS{{\it Mon. Not. R. Ast. Soc.} }
\def\NAT{{\it Nature} }

\def\PR{{\it Phys. Rev.} }
\def\PRL{{\it Phys. Rev. Lett.} }

\begin{document}

\titlepage
\bibliographystyle {unsrt}
\newcommand{\pa}{\partial}

\titlepage
\begin{flushright}
DF/IST--2.2000 \\
July  2000\\
\end{flushright}
\vspace{2cm}

\begin{center}
{\bf Fitting BOOMERANG and MAXIMA-1  data with a Einstein-Yang-Mills
  Cosmological Model}\\
\vspace{5mm}
{\grande }
\vspace{10mm}
M. C. Bento\footnote{ bento@sirius.ist.utl.pt}, O. Bertolami\footnote{ orfeu@cosmos.ist.utl.pt} and P. T. Silva

{\em Departamento de F\'{\i}sica,
  Instituto Superior T\'ecnico}\\ 

{\em Av. Rovisco Pais, 1049-001 Lisboa, Portugal } \\
\end{center}

\vspace{10mm}
\centerline{  Abstract}

\vspace{0.5cm}
\noindent
 
We analyse the implications of recent Cosmic Microwave Background (CMB) data for a specific cosmological model, based on the higher-dimensional Einstein-Yang-Mills system compactified on a $R\times S^3\times S^d$ topology and conclude that  the model parameters are tightly constrained by CMB spectra. 
Moreover, the model predicts  a relationship between the deceleration parameter at present,  $q_0$,  and some characteristic features of CMB spectra, namely the height of the first peak and the the location of the second peak, that is consistent with the observations and which can be further tested by future CMB and other experiments.

\vfill

%

\newpage
\setcounter{equation}{0}
\setcounter{page}{2}
Recent BOOMERANG data suggest that, within the framework  of inflation-motivated adiabatic Cold Dark Matter (CDM) models, the spatial curvature is close to flat and the primordial fluctuation spectrum is nearly scale invariant \cite{boom}. Moreover, the data prefer a baryon density $\Omega_b h^2$ somewhat higher than previous studies indicated, though consistent with  estimates from Big Bang nucleosynthesis.  When combined with large scale structure observations, this provides evidence for both dark matter and dark energy contributions to the total energy density $\Omega_{tot}$, in agreement with Supernova observations \cite{per} . On the other hand, MAXIMA-1, another balloon-borne experiment, finds new constraints on a seven-dimensional space of cosmological parameters within the class of inflationary adiabatic models \cite{max}:  $\Omega_{tot}=0.90\pm 0.15,~ \Omega_b h^2=0.025\pm 0.010,~ \Omega_{c} h^2=0.13\pm 0.10~(\Omega_c\equiv \Omega_{cdm}),~ n_s=0.99\pm 0.09$, all at the $95\%$ confidence level. By combining MAXIMA-1 results with high-redshift supernovae measurements, further constraints are obtained on the value of the cosmological constant and the fractional amount of matter in the Universe : $0.4 < \Omega_\Lambda < 0.76$ and $0.25 < \Omega_m < 0.50$.

In view of the increasingly tight constraints on cosmological parameters, fitting CMB  data with particle physics-motived cosmological models has become  a major goal for theoretical cosmology. Such models necessarily have to account for the negative pressure dark energy component of the energy density of the Universe, as revealed by  data. 
Although the  simplest and most obvious candidate for the missing energy is a vacuum
 density that contributes a fraction $\Omega_\Lambda 
\simeq 0.7$ of closure density, there are alternative, possibly theoretically advantageous candidate theories.
 Quintessence, as a substitute for the cosmological constant, is a slowly
 varying component with a negative equation of state.
 An example of quintessence is the energy associated with a scalar field
 slowly evolving down its potential \cite{cald} or a  scalar field coupled non-minimally with gravity \cite{uzan}.
Another possiblity for the nature of dark energy is the vacuum energy density of scalar and internal-space gauge fields arising from the process of dimensional reduction of higher dimensional gravity theories \cite{quinta}.

The purpose of this article is to fit CMB spectra, notably data recently obtained by the BOOMERANG and MAXIMA-1 experiments, within the framework of  a cosmological model based on the multidimensional Einstein-Yang-Mills system, compactified on a $R\times S^3 \times S^d$ topology \cite{quinta}. In particular, we analyse constraints coming from the measurement of the height and  positions of the first and second peaks in the CMB spectrum.


\vspace*{0.3cm}

The model is derived from the multidimensional Einstein-Yang-Mills-Inflaton  system

\begin{equation}
S = {1 \over 16 \pi  k} \int_{M^D} d^D {x}
\sqrt{- g}~\left[ R - 2  \Lambda+
{1 \over 8{e}^2} 
~{\rm Tr}  F_{\mu\nu} 
 F^{\mu\nu} 
- ~{1 \over 2} 
\left(\partial_{\mu}  \chi\right)^2 
-  U \left( \chi
\right)\right] 
\label{2.2}
\end{equation}
where $ g$ is $\det 
\left( g_{\mu\nu}\right)$, 
$ g_{\mu\nu}$ is the $D$-dimensional metric, 
$ R,~ F_{\mu\nu}\equiv F_{\mu\nu}^a \tau_a$, $ e$, $ k$ and $ \Lambda$ are, 
respectively, the scalar curvature, gauge field strength ($\tau^a$ being the generators of the gauge group that we assume to be SO(N), $N\geq d+3$), gauge coupling, 
 gravitational and cosmological constants 
in $D$ dimensions. We have included a scalar field, the inflaton  ($ \chi$),  with a potential  $ U(\chi)$,
responsible for the inflationary expansion of the external space and generation of the primordial energy density fluctuations.

After compactification  on a  $R\times S^3 \times S^d$ topology and setting the relevant fields to their vacuum configurations, the equations relevant for the resulting cosmological model are the following \cite{quinta}:

\begin{equation}
\left({\dot{a} \over a}\right)^2 = - {1 \over 4 a^2} 
+ {8 \pi k \over 3} \left[{\dot{\psi}^2 \over 2} + W(a, \psi)+\rho\right]~~,
\label{eq:2.17}
\end{equation}

\begin{equation}
\ddot{\psi} + 3 \left({\dot{a} \over a}\right) \dot{\psi}
+ {\partial W \over \partial \psi} = 0~~.
\label{eq:2.18}
\end{equation}
with 
\beq
W  =  e^{-d\beta\psi} \left[ -
{d(d-1)e^{-2\beta\psi} \over 64\pi k \vev{b}^{2}} + 
  {d(d-1)e^{-4\beta\psi} \over 8 e^2 \vev{b}^{4}} 
v_2 +  {\Lambda \over 8\pi k}\right]
+  e^{d\beta\psi} {3 \over 4 e^2 a^4} v_1~~,
\label{eq:2.19}
\eeq
where $\psi\equiv \beta^{-1} \ln \left( {b\over\vev{b} }\right) $, $a$ and $b$ being  the scale factors of $S^3$ and $S^d$, respectively, $e$ is the gauge coupling and $v_1,\ v_2={1\over 8}$ are the minima of the  potential, related with the external and internal components of the gauge fields.
The last term in Eq.~(\ref{eq:2.19}) represents the contribution
of radiation for the energy density of the Universe.
In Eq.~(\ref{eq:2.17}), $\rho$ denotes the energy density contribution of non-relativistic matter.

Different 
values for the cosmological constant $\Lambda$ correspond to different 
compactification scenarios (see Ref. \cite{bkm} and, for a quantum mechanical analysis, also Ref.~\cite{bert}). If $\Lambda>c_2/16\pi k$, where   
$c_2=[(d+2)^2(d-1)/(d+4)]e^2/16v_2$, 
there are no compactifying solutions and for
${c_1 \over 16\pi k}<\Lambda< {c_2 \over 16\pi k}$
with $c_1=d(d-1)e^2/16v_2$, a compactifying solution exists which is 
classically
stable but semiclassically unstable. Alternatively, if  
$\Lambda<c_1/16\pi k$, the effective 
4-dimensional cosmological constant, 
$\Lambda^{(4)}= 8\pi k W(a\rightarrow\infty,\psi)$, is negative.
Since  $\Lambda^{(4)}$ is required to  satisfy the order of magnitude observational bound
$\Lambda^{(4)} \approx 10^{-120}/ 16\pi k$,
we  fine-tune the multidimensional cosmological 
constant as
\beq
 \Lambda={c_1(1+\delta)\over 16\pi k}~,
\label{eq:lambda}
\eeq
 so that  $\delta$ is clearly 
proportional to $\Lambda^{(4)}$ and $c_1$ is determined by  choosing $\Lambda$ such that $\psi=0$ corresponds to the absolute minimum of the potential
in Eq.~(\ref{eq:2.19}), where  $\vev{b}^2 = {16\pi k v_2 /e^2}$. Hence 
\beq
\Lambda={d(d-1) \over 16\vev{b}^2}~(1 + \delta)~~.
\label{eq:2.22}
\eeq
Substituting Eq.~(\ref{eq:2.22}) into Eq.~(\ref{eq:2.19}), yields, 
in the large $a$ limit (implying that  the radiation term can be neglected)
\beq
W={d(d-1)\over 128 \pi k \vev{b}^2}\delta~~.
\label{eq:2.23}
\eeq

\vspace{0.3cm}
Notice that, although a non-vanishing  $\delta$ induces a semiclassical instability in the compactification solution, the decompactification time exceeds the age of the Universe by many orders of magnitude.

The deceleration parameter at present can be computed  differentiating
 Eq.~(\ref{eq:2.17}) and substituting the resulting term in 
$\ddot\psi$ by Eq.~(\ref{eq:2.18}) \cite{quinta}

\beq
q_0={-\delta_1 + {\epsilon\over 2}\over -{1\over 4 }+ \delta_1 + \epsilon}~~, 
\label{eq:2.27}
\eeq
where $\delta_1 \equiv d(d - 1) \alpha_{0} \delta_0 / 48$, $\alpha_0$ being  an order one constant  defined by $\left({a_0 \over \vev{b}}\right)^2 = \alpha_{0} 10^{120}$, $\delta=\delta_0~ 10^{-120}$ and
\beq
\epsilon  \equiv  {8\pi k\over 3}{\rho_0 a_0^2} = 
{3.2 \pi\over 3} \alpha_0\Omega_m h^2~~, 
\label{eq:2.28}
\eeq
where $0.4~\laq~h~\laq~0.7$
parametrizes the observational uncertainty in the Hubble constant, 
$H_0 = 100~h~km~s^{-1}~Mpc^{-1}$.

A  bound on $\delta_0$ can be obtained from $a_q\equiv a(t_q) =\alpha a_0$, where $t_q$ is the time  
when the vacuum contribution started dominating the dynamics of the Universe and $\alpha\equiv {a_q\over a_0}$ is a constant, equating the contributions of $W$ and $\rho(a_q)$ and using the  observational bound
 $\Omega_m~\laq~0.3$ \cite{bachall}. 
Thus, we get:

\beq
\alpha^3~\delta_0~\laq~{15.36~\pi \over d(d - 1)}~h^2~~,
\label{eq:2.29}
\eeq   
Since the red-shift of the supernovae data indicating the accelerated 
expansion of the Universe is $z \ge 0.35$, then  
$\alpha\le 0.74$ and, for $d = 7$ and $h= 0.5$, we obtain

\beq
\delta_0~\laq~0.71~,
\label{eq:2.30}
\eeq
which implies, for e.g.  $\delta_0 = 0.7, \alpha_0=5$, that

\beq
q_0 = -~0.59~,
\label{eq:2.31}
\eeq
 Note that we have corrected numerical values in  Eqs.~(\ref{eq:2.28}) and (\ref{eq:2.29}) of
 Ref.~\cite{quinta}, but the main result, Eq.~(\ref{eq:2.31}), remains within
 the   most likely region of values for $q_0$, as revealed by
observational data \cite{per}.

\begin{figure}
\centerline{\epsfysize=11cm \epsfbox{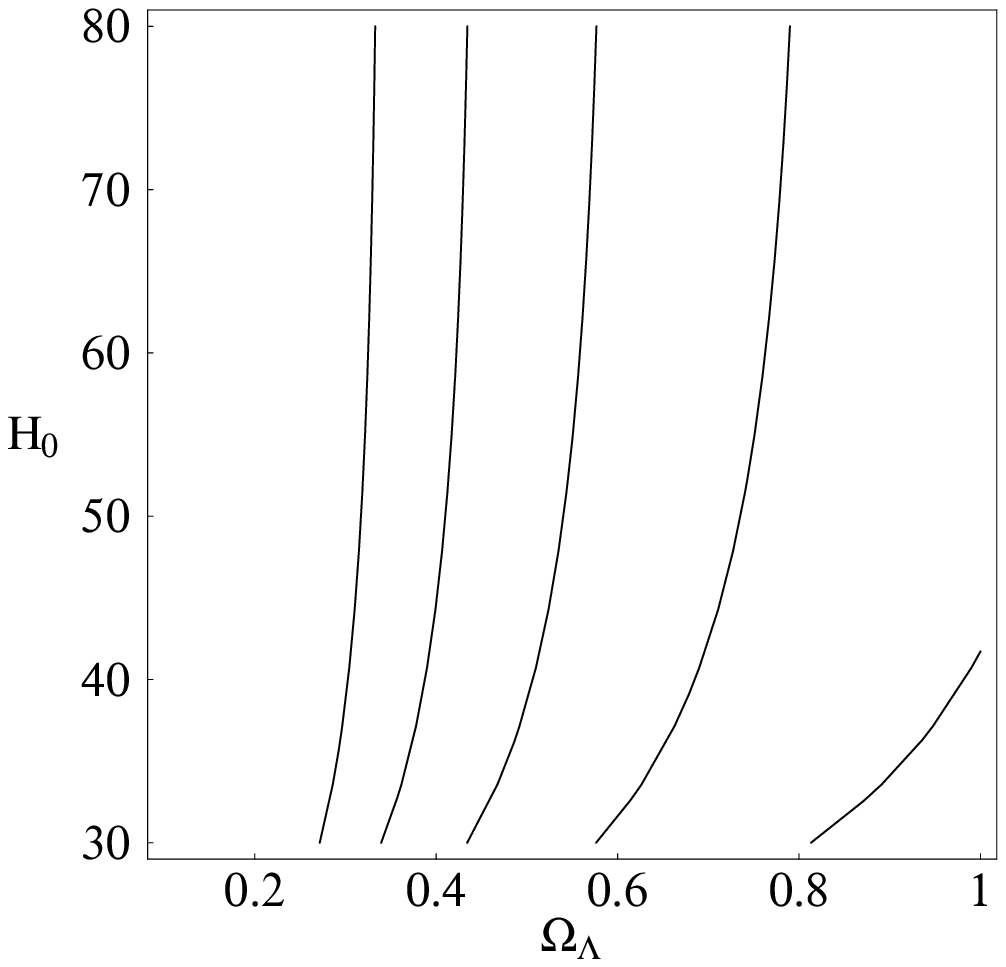}}
\caption{Contours of $q_0$, from $-0.3$ (right)  to $-0.7$ (left),  at $0.1$ intervals, in the two-dimensional parameter space $\Omega_\Lambda, H_0$  (in units km/sec/Mpc), for $d = 7, \Omega_m = 0.3$ and $\alpha_0 = 5$, in  our compactified Einstein-Yang-Mills system   model.}
\label{fig:q0delh}
\end{figure}

We show, in Figure~\ref{fig:q0delh},  contours of $q_0$ in the two-dimensional parameter space $(\Omega_\Lambda,~ H_0)$, $\Omega_{\Lambda} $ being the vacuum energy density, in which it varies most strongly. Indeed, we have checked that variation with $\Omega_m$ and $\alpha_0$ is modest.

A distinct feature of the model is that, in spite of having a 
closed topology,  a phase of accelerated 
expansion can take place \cite{quinta}. We would like to stress that (slightly) closed models are actually favoured by  recent BOOMERANG data. Very closed models work   by increasing  $n_s$, the scalar spectral index, and $r$, the gravity waves contribution \cite{zalda}.

\begin{figure}
\centerline{\epsfysize=11cm \epsfbox{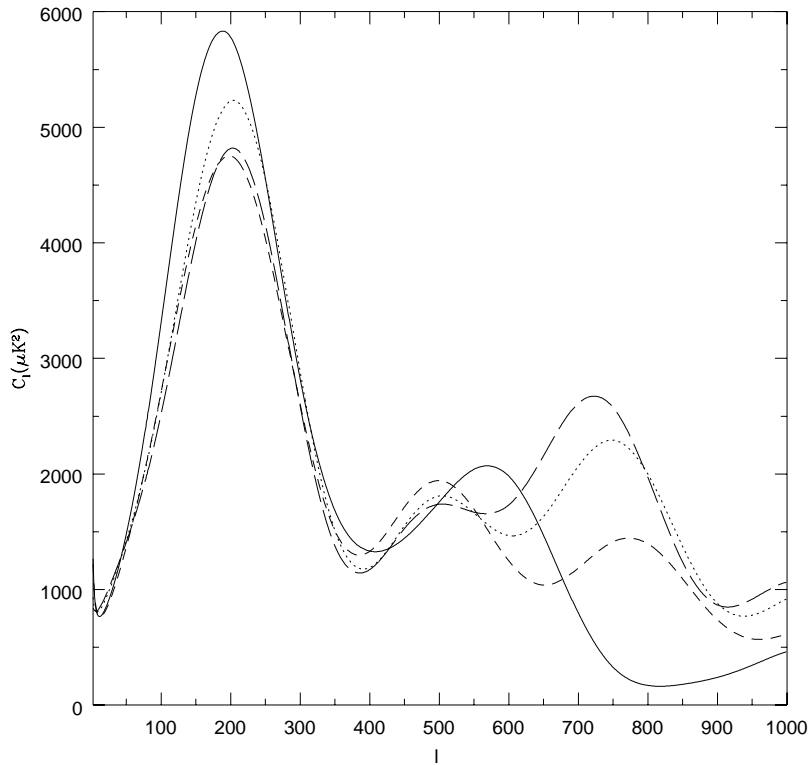}}
\caption{CMB angular power spectra, $C_l=l(l+1)\vev{\vert T_{lm}\vert^2}/(2\pi)$, where the $T_{lm}$ are the multipole moments of the CMB spectra. The three curves correspond, respectively, to Models 1, 2  and 3, defined by specific values of  $\Omega_{tot}, \omega_b, \omega_c, \Omega_\Lambda, n_s, \tau_C$ (see text), considered to be good fits to B98+COBE data \cite{boom}.}
\label{fig:specboom}
\end{figure}

In Figure~\ref{fig:specboom}, we show the CMB power spectra that correspond to the four sample CDM models that fit B98+COBE  data. These are best-fit theoretical models defined by the value of six parameters  using sucessively more restrictive ``prior probabilities'' on the parameters \cite{boom}. The  parameters are:   ($\Omega_{tot}, \omega_b, \omega_c, \Omega_\Lambda, n_s, \tau_C$),  where $\omega_{b,c}=\Omega_{b,c} h^2$ are  the cosmological baryon and CDM densities,   $\tau_C$ is the optical depth to Thompson scattering from the epoch at which the Universe reionized to the present. Hence, $ \Omega_{tot}\equiv \Omega_b + \Omega_c + \Omega_\Lambda=1-\Omega_k$ where $\Omega_k$ is the curvature density. Model 1 fixes these parameters at $(1.3, 0.10, 0.80, 0.6, 0.80, 0.025)$, Model 2 at ($1.15, 0.03, 0.17, 0.44, 0.925, 0$), Model 3 at ($1.05, 0.02, 0.06, 0.90, 0.825, 0$) and Model 4 at
 ($1.0, 0.03, 0.27, 0.60$, $ 0.975, 0$). For $d=7$, these models fix parameter $\delta_0$ in our model as :
$1.88, 0.26, 1.17$ and $1.096 $, respectively. On the other hand, our EYM-motivated scenario predicts the corresponding $q_0$ values to be, respectively,  $-0.03, 0.13, $ $-0.71  $ and $-0.23 $, for $\alpha_0=5$; hence,  model 2 would be excluded in this  scenario, for reasonable values of $\alpha_0$.

Next, we study the dependence of $q_0$  on certain characteristic features of CMB spectra which are being increasingly constrained by CMB experiments, mainly the height of the first acoustic  peak, $A_1$, its position, $l_1$ and the location of the second peak, $l_2$.

The height of the primary peak  is controlled mainly by the  baryon-to-photon ratio, varying as $\Omega_b h^2$,  the dark matter-to-photon ratio, varying as $\Omega_c h^2$ and the cosmological constant \cite{linew}. In fact, $\Omega_b h^2$ and $\Omega_c h^2$  produce competing effects: when $\Omega_b h^2$ increases $A_1$ increases and when $\Omega_c h^2$ increases, $A_1$ decreases. 
The effect of varying $\Lambda$, holding  $\Omega_b h^2$ and $h$  fixed, is that the largest values of $\Lambda$ correspond to the largest Doppler peaks. On the other hand, as $h$ increases, $A_1$ decreases and $l_1$ shifts to larger scales. Observations indicate that $l_1\simeq 200$.
The spectral index $n_s$  also changes the amplitude of the peaks such that, as $n_s$ increases, $A_1$ also increases. The low second peak found by BOOMERANG and MAXIMA-1 can be fit by either decreasing the tilt $n_s$ or by increasing the baryon density compared to the usually assumed values $n_s\approx 1,\ \omega_b\approx 0.02$, although both of these solutions have problems of their own \cite{zalda}.

The location of the second peak in the CMB power spectrum depends mainly, if the geometry is fixed, on the expansion rate of the Universe at the epoch of recombination \cite{hu}, and this depends on the nonrelativistic matter density and the Hubble constant.  Of course, the precise location of the second peak can change upon variations in several other parameters; however, it changes  very little as each of these  parameters is allowed to vary within its acceptable range \cite{kamio}.
 We consider $\Omega_b h^2$ in the  range advocated in Ref.~\cite{tyler}, from measurements of the deuterium abundance,  $\Omega_b h^2= 0.019 \pm 0.001$. 
On the other hand,  allowable variations in $n_s$ ($n_s=0.99\pm 0.010$ \cite{max}) lead to even smaller uncertainties in the second-peak location than those from uncertainty in the baryon density.

\begin{figure}
\centerline{\epsfysize=11cm \epsfbox{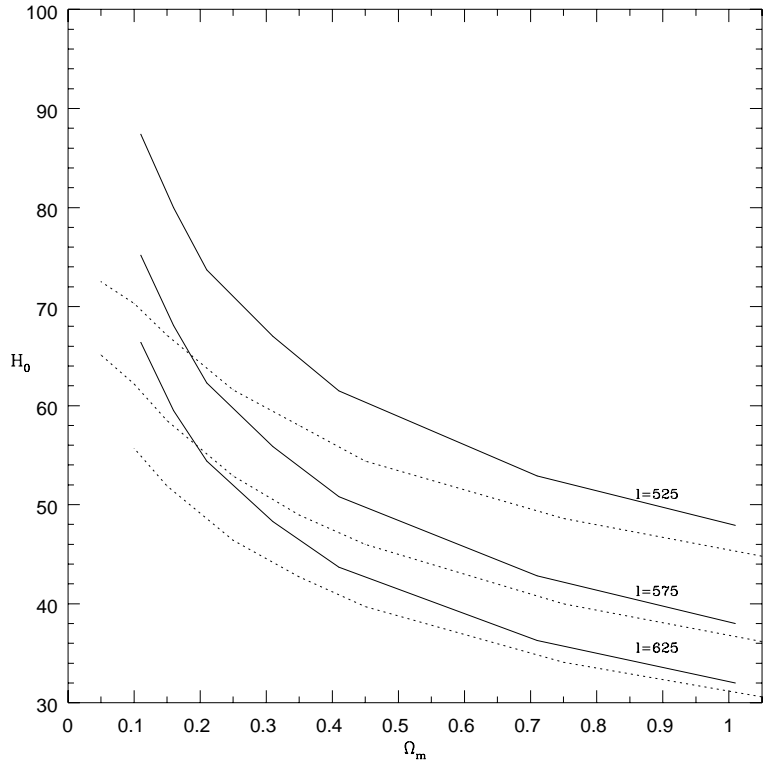}}
\caption{ Contours of $l_2$, the multipole moment at which the second peak in the CMB power spectrum occurs, in the two-dimensional parameter space  $\Omega_m, H_0$ (in units km/sec/Mpc), for  $\Omega_b h^2=0.019$, in two slightly closed models $\Omega_{tot}=1.01$ (full curve),~$1.05$ (dashed).}
\label{fig:cl2}
\end{figure}

Figure~\ref{fig:cl2} shows contours of $l_2$, the multipole moment at which the second peak in the CMB power spectrum occurs, in the two-dimensional parameter space  $(\Omega_m, H_0)$, in which it varies most strongly, for $\Omega_b h^2=0.019$, in two slightly closed models $\Omega_{tot}=1.01, 1.05$ (B98 data with medium priors suggests $ 0.88<\Omega_{tot}<1.12$ at $95\%$ confidence level).

\begin{figure}
\centerline{\epsfysize=11cm \epsfbox{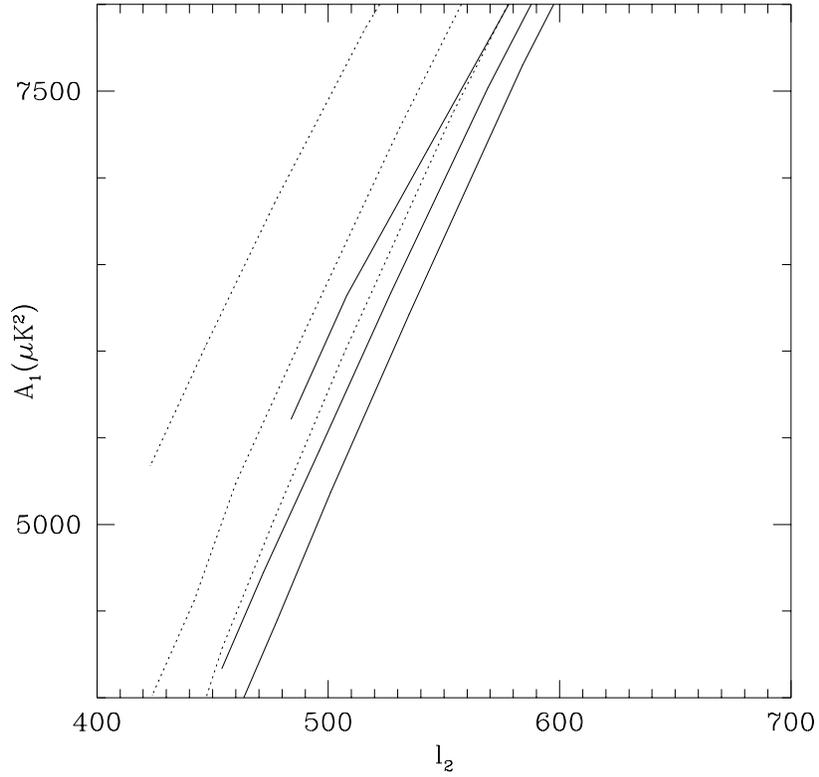}}
\caption{Contourplots of $q_0=-0.25$ (bottom),$ -0.5$ and $-0.75$ (top) in the $( l_2, A_1)$ parameter space, for $\Omega_{tot}=1.01$ (full curve), $1.05$ (dashed) and  $\Omega_b h^2=0.019$ $(\alpha_0=5)$.}
\label{fig:q0ot}
\end{figure}

In Figure~\ref{fig:q0ot}, we show contourplots of $q_0=-0.25, -0.5, -0.75$ in the $(A_1, l_2)$ parameter space, for $\Omega_{tot}=1.01, 1.05$ and $\Omega_b h^2=0.019$. Once these latter parameters are fixed, we see that our Einstein-Yang-Mills model can be greatly constrained by CMB data as  knowledge of the values $(A_1, l_2)$ determines $q_0$. We find that, for $l_2\simeq 500$, our results do not vary significantly for different $\Omega_b h^2$ values. In turn, once the value of $q_0$ is known, the relevant parameters of our model, e.g. $\alpha_0$ and $\delta_0$, become fixed (see Figure~\ref{fig:q0delh}). Furthermore, as expected,  only for flat or slightly closed models the observational constraints on $A_1~(4500\mu K^2~\laq~ A_1~\laq~ 5500\mu K^2)$ and $l_2~(450~\laq~l_2~\laq~ 600)$  can be satisfied; indeed, already for $\Omega_{tot}=1.05$ the corresponding curves are broken, meaning that observational constraints  cannot be met  for the full range of $l_2$, $A_1$ values considered, and the situation worsens as $\Omega_{tot}$ increases.


We conclude that recent CMB data tightly constrain our Einstein-Yang-Mills-inspired cosmological model. The strongest prediction of the model is the connection that can be inferred between the acceleration parameter and the height of the first peak together with the  location of the second peak in the CMB spectrum. This connection can be further tested in the near future, as upcoming CMB data put  tighter constraints on $A_1,~l_2$ and $q_0$ becomes also more constrained via supernovae observations and   other experiments (see \cite{bachall} and references therein). In particular, we find that our model is consistent with available CMB data, favouring flat or very slightly closed models and $q_0\simeq -0.5$.
\bs

We have used CMBFAST \cite{cmbfast} to calculate the CMB power spectra.

\vspace*{0.3cm}


\newpage

\newpage

\newpage

\newpage

\end{document}